\documentclass[10pt,conference]{IEEEtran}

\usepackage[normalem]{ulem}
\usepackage{cite}
\usepackage[keeplastbox]{flushend}
\usepackage{color,soul}
\usepackage{amsmath,amssymb,amsfonts}
\usepackage{multirow}
\usepackage{graphicx}
\usepackage{float}
\usepackage{todonotes}
\usepackage{array}
\usepackage{booktabs}
\usepackage{amsmath}
\usepackage{algorithm}
\usepackage[noend]{algorithmic}
\usepackage{tabularx}
\usepackage{textcomp}
\usepackage{mathtools}
\usepackage{xcolor}
\usepackage{bm}
\usepackage{tablefootnote}
\usepackage{verbatim}
\usepackage[hyphens]{url}

\def\BibTeX{{\rm B\kern-.05em{\sc i\kern-.025em b}\kern-.08em
    T\kern-.1667em\lower.7ex\hbox{E}\kern-.125emX}}
\usepackage[bookmarks=true,breaklinks=true,letterpaper=true,colorlinks,citecolor=blue,linkcolor=blue,urlcolor=blue]{hyperref}
\DeclarePairedDelimiter\ceil{\lceil}{\rceil}
\DeclarePairedDelimiter\floor{\lfloor}{\rfloor}
\newcolumntype{Y}{>{\centering\arraybackslash}X}

\usepackage[all=normal, floats, mathspacing,wordspacing,leading, paragraphs, charwidths, lists]{savetrees}
\title{PhotoFourier: A Photonic Joint Transform Correlator-Based Neural Network Accelerator}
\author{   \IEEEauthorblockN{Shurui Li\IEEEauthorrefmark{1}, Hangbo Yang\IEEEauthorrefmark{1}, Chee Wei Wong\IEEEauthorrefmark{1}, Volker J. Sorger\IEEEauthorrefmark{2}, Puneet Gupta\IEEEauthorrefmark{1}}
    \IEEEauthorblockA{\IEEEauthorrefmark{1}University of California, Los Angeles
    \\\{shuruili, yanghumble, cheewei.wong, puneetg\}@ucla.edu}
    \IEEEauthorblockA{\IEEEauthorrefmark{2}The George Washington University
    \\\{sorger\}@email.gwu.edu}}

\begin{document}
\maketitle
\thispagestyle{plain}
\pagestyle{plain}


\begin{abstract}
The last few years have seen a lot of work to address the challenge of low-latency and high-throughput convolutional neural network inference. Integrated photonics has the potential to dramatically accelerate neural networks because of its low-latency nature. Combined with the concept of Joint Transform Correlator (JTC), the computationally expensive convolution functions can be computed instantaneously (time of flight of light) with almost no cost. This `free' convolution computation provides the theoretical basis of the proposed PhotoFourier JTC-based CNN accelerator. PhotoFourier addresses a myriad of challenges posed by on-chip photonic computing in the Fourier domain including 1D lenses and high-cost optoelectronic conversions.
The proposed PhotoFourier accelerator achieves more than $28\times$ better energy-delay product compared to state-of-art photonic neural network accelerators. 

\end{abstract}
\section{Introduction} \label{sec:introduction}
Convolutional neural networks (CNNs) play a key role in modern Artificial Intelligence (AI) technologies and are the core of many computer vision applications including image classification \cite{alexnet,vgg16,resnet}, object tracking \cite{yolo,rcnn}, medical imaging \cite{medicalcnn1,medicalcnn2}, etc. Over the past decade, there have been many efforts of designing domain-specific accelerators utilizing parallel architectures to accelerate the computation of neural networks in an energy-efficient way \cite{eyeriss, diannao, dadiannao, systolicarray1, pragmatic}. However, the rapidly growing size of modern CNNs and the slowdown of Moore's law have limited CMOS digital accelerators in terms of the energy cost of data movement and computation \cite{eyeriss, drisa}. Silicon photonics has emerged as a promising approach to deliver massive compute parallelism and high efficiency \cite{shiflett2021albireo,liu2019holylight,shiflett2020pixel}. Photonic components can easily operate above 10 GHz while still being relatively low-power \cite{albireomrr, advancedmrr}, and photonic waveguides do not suffer from RC delay or energy losses \cite{opticalinterconnect1, gf45nm}. These features give photonics an unmatched advantage in low-latency and low-power computation. 

Photonic neural network accelerators can be roughly classified into two main categories: Mach-Zehnder Interferometer (MZI) and micro-ring resonator (MRR) based dot product accelerators \cite{shiflett2020pixel, shiflett2021albireo, DEAPCNN2019,  onchip_ONN2018, pcnna, liu2019holylight,zokaee2020lightbulb, jiaqionchipfft, crosslight} and Fourier optics-based convolution accelerator \cite{standford4f, miscuglio2020optica, 4farchitecurspie}. Most MZI/MRR dot product accelerators resemble compute-in-memory analog accelerators \cite{analogaccelerator1,inmem1,inmem2}, but with high clock frequencies (5-10 GHz). The large number of large-sized MZIs and/or MRRs required can become a problem. On the other hand, Fourier optics-based designs typically utilize the convolution theorem to accelerate the convolution operation, which states that convolution in the space domain is equivalent to point-wise multiplication in the Fourier domain. Such systems, typically called 4F systems (total system length is 4 times the focal length of the lens), leverage time-of-flight (and passive, hence, zero energy) Fourier transform using Fourier lenses to reduce the complexity of convolution from $O(N^2)$ to just $O(N)$. A point-wise multiplication unit is required at the Fourier plane (after the Fourier transform) and the filter weights are directly loaded into the multiplication unit \cite{miscuglio2020optica}.
Theoretically, compared to dot product accelerators, 4F systems can perform the same computation with significantly fewer optical components because of the complexity reduction. 
However, 4F systems require Fourier domain filters that are complex-valued, with sizes same as inputs. This constraint makes 4F systems harder to implement as supporting complex multiplication is hard. Moreover, it makes 4F systems less efficient when executing conventional CNNs, which typically use $3 \times 3$ real-valued filters. All prior works on 4F-based CNN accelerator are prototypes using free-space optics \cite{miscuglio2020optica, standford4f}, which are slow and bulky compared to on-chip photonics. 

In this work, we propose using Joint Transform Correlator (JTC) to accelerate CNNs by reducing the computation complexity through Fourier optics, while addressing the issues faced by typical 4F systems. JTC is a variant of Fourier optics that computes the auto-convolution of two input signals using a pair of Fourier lenses. Just like 4F systems, JTC also takes the advantage of the `free' Fourier transform but uses spatial filters instead of complex-valued Fourier filters. Therefore, JTC systems allow filters to be smaller than inputs and only need to support real-valued multiplication.

In this paper, we present PhotoFourier, a photonic CNN accelerator based on Joint Transform Correlator (JTC). The main contributions can be summarized as follows:
\begin{itemize}
      \item We propose the row tiling/partitioning algorithm to implement 2D convolutions using 1D on-chip lenses.
    \item We develop a temporal accumulation approach to cut down Analog-to-Digital Converter (ADC) power by 16X and improve neural network accuracy significantly.
     \item To the best of our knowledge, this is the first work to propose the architecture design of an on-chip Fourier-optics based photonic neural network accelerator. PhotoFourier can achieve as much as $28\times$ better energy-delay product compared to state-of-art photonic neural network accelerators.
\end{itemize}

\section{A Primer on the JTC system} \label{sec: jtcprimer}
\subsection{Background of JTC}

JTC has been widely used for many applications including optical encryption \cite{jtcencrypt1,jtcencrypt2, jtcencrypt3}, image filtering \cite{jtcfiltering1,jtcfiltering2}, and object tracking \cite{jtctracking1,jtctracking2,jtctracking3} over the past two decades. 
Recently there has been a growing interest in optical and photonic neural networks, with some works trying to realize JTC-based optical neural networks. \cite{george2022JTCarxiv, jtccitation2,jtccitation3,jtccitation4} provides theoretical analysis and experimental demonstration of a free-space JTC system designed for low latency convolution operations while \cite{jtcpresentation} demonstrates the concept of a basic on-chip JTC-based photonic neural network.

\begin{figure}[h]
    \centering
    \includegraphics[width=0.95\columnwidth]{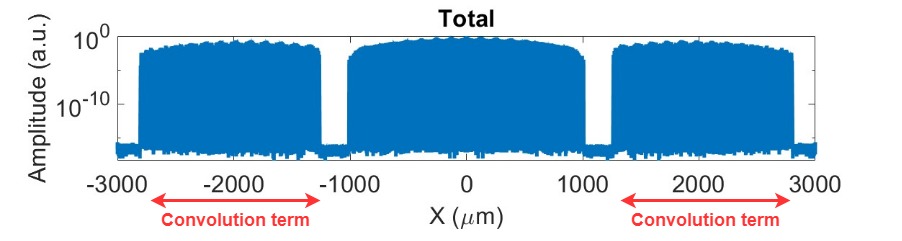}
    \caption{Simulated JTC output for a 256-element input (partitioned from a CIFAR-10 input) with tiled convolution kernels.}
    \label{fig:measurementplot}
\end{figure}

In physics, an optical lens can achieve Fourier transform $\mathcal{F}[\tilde{E}(x, y, f)]$ \cite{jtccitation6} on its back focal plane if an input image $\tilde{E}(x, y, f)$ illuminated by a coherent light (usually a laser) is at the front focal plane of the lens. $\tilde{E}$ is the amplitude of the light at the front focal plane, $\mathcal{F}$ is the symbol of the Fourier transform. Adopting the Fourier transform of the lens, \cite{weaverjtc1966} first made an optical JTC to generate the optical convolution with both phase and amplitude. Based on the traditional 2D optical JTC, a baseline 1D on-chip photonic JTC can be built with slight modifications. Figure \ref{fig:pichighlevel} (a) depicts the layout of a baseline on-chip JTC system, which consists of five key components: (1) a 1D multi-channel input beam with a signal $s(x+x_s)$ and a kernel $k(x-x_k)$ (where $x_s$ and $x_k$ are offsets of $s$ and $k$ from the global origin in $x$ direction, respectively) passes through (2) the first on-chip metasurface-base lens functioning as a traditional free-space lens, to achieve 1D Fourier transform $\mathcal{F}\left[s\left(x+x_{s}\right)+k\left(x-x_{k}\right)\right]$, (3) a \emph{nonlinear function} component implemented using photodetectors (to transfer optical signals to electrical signals meanwhile achieve a \emph{square function}) and electro-optic modulators (EOM) (to transfer electrical signals back to optical signals), (4) the second on-chip metasurface-base lens, and arrives at (5) photodetectors recording the intensity pattern of the convolution computed by the JTC:
\begin{equation} \label{eqn:jrcoutput}
    s(x+x_s + x_k) * k(-x) + s(-x) * k(x - x_s - x_k) + O(x) 
\end{equation}
, where $*$ means convolution, $O(x) = \mathcal{F}\left[\left|S\left(x\right)\right|^{2}+\left|K\left(x\right)\right|^{2}\right]$. The first and second terms are the computed auto-convolution between the two inputs whereas the third term $O(x)$ is a non-convolution term.
The convolution terms in Equation \ref{eqn:jrcoutput} can be shifted off the center non-convolution term $O(x)$ by adjusting the distance between two inputs, so that the convolution would not be affected by the non-convolution term $O(x)$. The photodetectors only need to detect one of the convolution terms to get the convolution result. To demonstrate this, we simulate the JTC output of a 256-element input which is a partitioned and tiled CIFAR-10 input, with a tiled convolution kernel (refer to Section \ref{sec:convapproxmethod} for tiling details), and the JTC output is shown in Figure \ref{fig:measurementplot}. The simulated output clearly shows the three terms in the output are spatially separated with no overlap. 

In the baseline system, the non-linear function is a square function achieved by photodetectors. One photodetector and one MRR are required for each waveguide. EOMs in this design are tunable MRRs\cite{albireomrr}, which transfer electrical signals back to optical signals. MRRs that implement the square function can be directly controlled by the output of the photodetectors, without conversion between analog and digital domains.
The distinction between on-chip JTC and conventional free-space JTC is that 2D lenses are replaced with 1D on-chip lenses, hence 1D convolutions are computed instead of 2D convolutions.
 
\subsection{A JTC accelerator prototype} \label{sec: baselinejtc}
\begin{figure}[h]
    \centering
    \includegraphics[width=1.0\columnwidth]{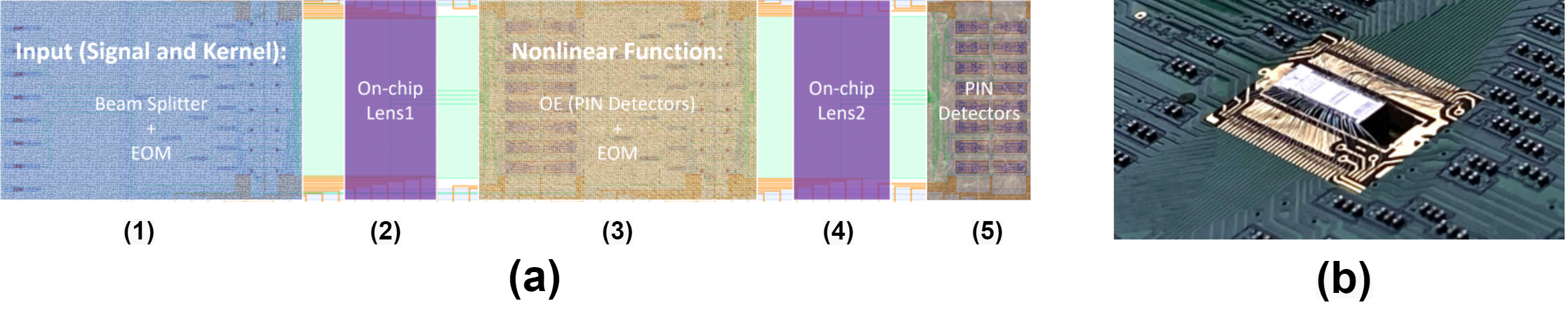}
    \caption{(a): The annotated layout diagram of a baseline on-chip JTC system. (b): The PCB photo of the fabricated prototype.}
    \label{fig:pichighlevel}
\end{figure}

We have designed and fabricated a prototype of the baseline system, which is the first on-chip JTC system. Figure \ref{fig:pichighlevel} (b) shows the fabricated JTC chip inside a PCB. 
The detailed experimental evaluation of the prototype system is out of the scope of this paper, as we focus on the architecture design and analysis of an \emph{upscaled} system. Still, the prototype system demonstrates that on-chip JTC systems are suitable and realizable in terms of accelerating CNNs.

\subsection{Issues faced by on-chip JTC accelerators}
The advantage of reducing the complexity of convolution operation without adding weight bandwidth overhead makes JTC a potentially better candidate than other photonic systems for efficiently accelerating CNNs. However, there are still many challenges that need to be addressed. Some issues are faced by photonic accelerators in general while the others are specific to on-chip JTC accelerators.

\subsubsection{1D lens}
Being on-chip means the lenses can only be one-dimensional, hence only 1D Fourier transform is supported and results in 1D convolution. Most CNNs use 2D convolution to capture information on both x and y dimensions. Clearly, just using 1D convolution will lead to poor accuracy and make JTC systems not able to execute conventional CNNs. To overcome this challenge, we propose the row tiling method to approximate 2D convolutions with 1D convolutions accurately. 

\subsubsection{Component redundancy}
The baseline JTC system described in Section \ref{sec: baselinejtc} can be split into two identical parts. Each part contains a set of MRRs, Fourier lens, and photodetectors. When processing a convolution, both parts can not be utilized at the same time, resulting in a 50\% utilization. Such inefficiency leads to potential optimizations including pipelining the system, which will be discussed in Section \ref{sec:pfcu}.

\subsubsection{Overhead of the non-linear function implementation}
A baseline JTC system uses MRRs to implement the required non-linear function, which results in undesired power and area overhead. In fact, the non-linear function could be implemented passively using optical non-linear materials, which can massively reduce the total number of active photonic components. Promising research results have been reported on optical non-linear materials \cite{nonlinearmaterial_sorgergroup, nonlinearmaterial2015, nonlinearmaterial2016alam} and JTC systems with non-linear materials \cite{nonlinearjtc1994, nonlinearjtc2020encoding, george2022JTCarxiv}. The reason for not using non-linear materials in the baseline JTC system is that such materials are not mature enough to be fabricated with silicon photonics. However, in the near future, passive non-linear materials could be used to implement the non-linear function, making designs more power efficient. 

\subsubsection{O-E and E-O conversion overhead}
Theoretically, the power efficiency of photonic accelerators should be an advantage over digital accelerators but the components required for O-E and E-O conversions (ADC, DAC, and modulators) are active and draw a large amount of power. If the architecture is not carefully designed to compensate for the conversion overhead, the overall power efficiency of photonic accelerators can easily be worse than CMOS accelerators. A key part of our architecture and dataflow design is to minimize the number of O-E and E-O conversions for optimal power efficiency.

\subsubsection{Mismatch between the frequency of photonics and CMOS} \label{sec:freqmismatch}
One advantage of silicon photonics is that they can be clocked extremely fast. Optical components like MRRs can operate above 30 GHz. Most existing photonic neural network accelerators set clock frequency between 5 to 10 GHz. However, it is extremely challenging to design and fabricate CMOS components with 10 GHz frequency. CMOS circuit is required to generate inputs, receive outputs, communicate with memory, and compute operations that the photonic accelerator is not able to compute. How to address the frequency mismatch between CMOS circuits and photonics is a challenge toward realistic photonic neural network accelerator designs. 

\section{2D Convolution Computation on JTC} \label{sec:convapproxmethod}
As discussed in Section \ref{sec: jtcprimer}, an on-chip JTC can compute the convolution between two inputs, but is limited to 1D.
To address this issue, we propose a generic algorithm to compute 2D convolution using 1D convolution, which can be applied to any hardware that supports 1D convolution, including JTC systems.
The key idea of the proposed algorithm is row tiling (and partitioning), where the rows of 2D inputs and kernels are tiled to form 1D inputs and kernels for 1D convolution. The proposed algorithm can achieve identical results as 2D convolutions in `valid' mode (without zero padding, output size smaller than input size), and can closely approximate 2D convolutions in `same' mode (with zero padding, output size same as input size). In the rest analysis, we assume 2D convolution uses the `same' mode, which is more common.  Assuming a 2D input has size $S_i \times S_i$, a 2D kernel has size $S_k \times S_k$, and the maximum 1D convolution size supported $N_{conv}$. Depending on $S_i, S_k$, and $N_{conv}$, the algorithm is split into three variations. 
\begin{figure*}[h]
    \centering
    \includegraphics[width=1.0\linewidth]{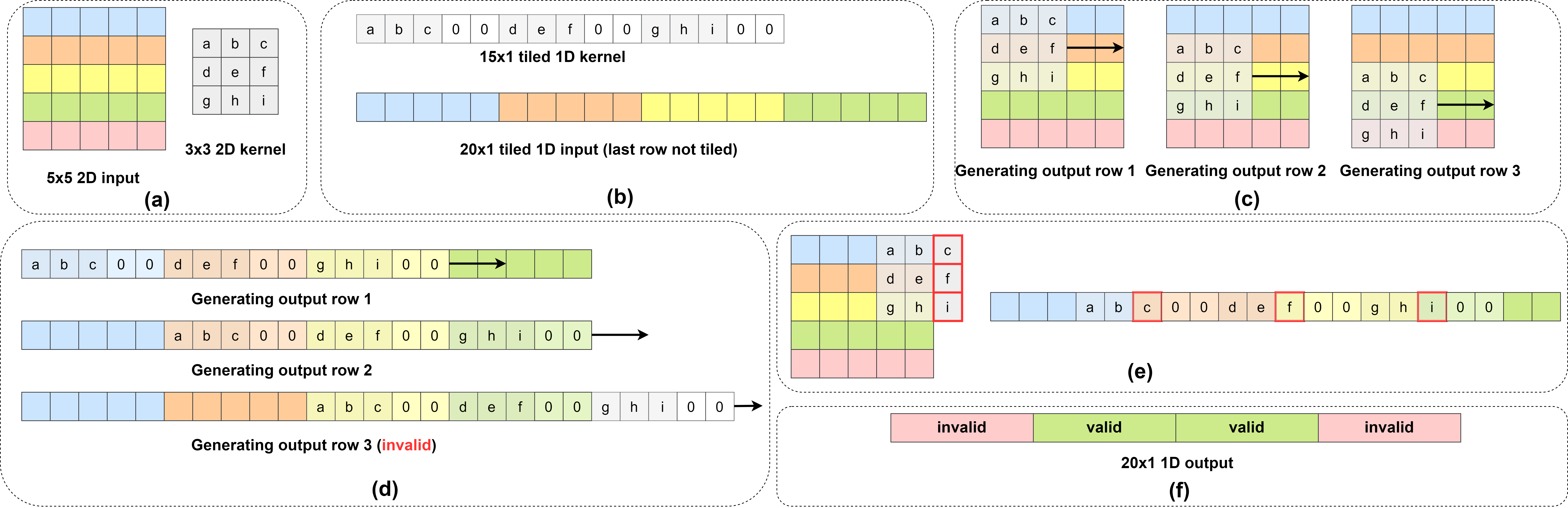}
    \caption{Visualization of row tiling with an example of $5 \times 5$ input, $3 \times 3$ kernel, and maximum 1D convolution size of 20. Different rows of the input are represented using different colors. (a): 2D input and kernel. (b): Tiled 1D input and kernel. Kernel rows are zero-padded to match the input row size. The last row of input is not tiled due to the limit of 1D convolution size. (c): Sliding window convolution process of normal 2D convolution to produce rows 1-3 of the output. (d): Sliding window convolution process of 1D convolution. For the first two rows, the tiled kernel rows are aligned with their corresponding input rows and produce valid results. Row 3 illustrates the case where the tiled kernel `slides' outside the input, and generates invalid results (since input row 5 is not tiled). (e): Edge effect. For 2D convolution with zero-padding when the filter is sliding outside of the inputs, the part outside the input ($c,f,j$) will convolve with zero. However, for 1D convolution they will convolve with the next input row, producing different results compared to 2D convolution. (f): Output format of 1D convolution. Invalid results are marked with red color.}
    \label{fig:tilingvisualizationfull}
 
\end{figure*}

\subsection{Row tiling}
\label{sec:rowtiling}
Row tiling can be applied when $N_{conv} > S_k \times S_i$, which is the most common case. The process is best explained with the visualizations of sliding window convolution, which are shown in Figure \ref{fig:tilingvisualizationfull}.
The first step is to tile the rows of the 2D input and kernel. The number of input rows that can be tiled each time is $\floor*{\frac{N_{conv}}{s_i}}$, which depends on the maximum size of 1D convolution. All kernel rows are tiled, but they are separated by zero-padding with size $S_i - S_k$ to ensure input and kernel rows are aligned after tiling (Figure \ref{fig:tilingvisualizationfull} (b)). Zeros are added to the end of tiled input and kernel rows, to make both of them have length $N_{conv}$. For conventional CNNs, the 2D convolution process can be visualized by sliding the kernel over the input, and in each step the overlapped regions between the kernel and input are multiplied and accumulated (dot product) to generate a single output value. This step is repeated until the kernel is convolved with the entire input (Figure \ref{fig:tilingvisualizationfull} (c)). Similarly, for 1D convolution, the 1D kernel is sliding from left to right and the dot product is computed for the overlapped region. Since kernel rows and input rows are aligned after tiling, 1D convolutions essentially perform the same computation as 2D convolutions and generate the same results (first two rows of Figure \ref{fig:tilingvisualizationfull} (d)).  
The outputs are valid 2D convolution results as long as the tiled kernel is fully inside the tiled input rows. However, when continuously sliding the 1D kernel as shown in the last row of Figure \ref{fig:tilingvisualizationfull} (d), the dot product results are invalid, since filter row 3 ($g,h,i$) is not convolving with the correct input row (row 5 of original input cannot be tiled).  For the example in Figure \ref{fig:tilingvisualizationfull}, a 20-element output is generated, but only the middle 10 elements are valid convolution results (two valid output rows). For the cases where the entire 2D input cannot be fully tiled, the tiling will be repeated until all the valid output rows are generated. 
The general formula for the number of valid output rows $N_{or}$ that can be generated through row tiling in one convolution operation is $$N_{or} = \floor*{\frac{N_{conv}}{s_i}} - S_k + 1$$, and the total number of 1D convolution required is $\ceil*{\frac{S_i}{N_{or}}}$.
Therefore the computation efficiency (measured by the percentage of valid outputs) is higher when $N_{conv}$ is large or $S_i$ is small.

\emph{Edge effect:} 2D convolution in `same' mode pads input edges with zero. The output of the proposed row tiling algorithm will be different in the regions where a single kernel row overlaps with two input rows because row tiling does not pad inputs (Figure \ref{fig:tilingvisualizationfull} (e)). 
The difference only happens at the edges of original input rows and the impact is minimal, especially for small kernels. Zero-padding can be applied during tiling so that the proposed algorithm can generate identical results compared to 2D convolution. However, adding zero-padding will make the output size larger than the input, which leads to additional overheads caused by extracting the desired output. Since the impact of the edge effect is small (Section \ref{sec:rowtilingaccuracy}), zero-padding is not applied by default.

\subsection{Partial row tiling}
\label{sec:partialrowtiling}
When $S_i <= N_{conv} < S_k \times S_i$, not enough input rows can be tiled to generate an entire row of 2D convolution output in one step. In this case, tiling can still be applied but multiple cycles are required to obtain the full results of one output row. 

For example, when $N_{WA} = 2 \times S_A$, the computation of a single output row is split into two cycles and the results are accumulated after both cycles complete the execution. In cycle 1 the first two rows of the input and the kernel are tiled while in cycle 2 only the third row of input and kernel are processed (under-utilizing the convolution hardware). Number of cycles required to compute a full 2D convolution is $S_i \times \ceil*{\frac{S_k}{N_{ir}}}$, where $N_{ir} = \floor*{\frac{N_{conv}}{S_i}}$ (the number of input rows can be tiled).

\subsection{Row partitioning}
When $N_{conv} < S_i$, a single row of input needs to be split into multiple partitions. The partitioning is similar to the case where $N_{conv} = S_i$ (dividing the 2D input into individual rows), except that each input row is further divided into partitions. The total number of cycles required to compute a full 2D output plane is $S_i \times S_k \times \ceil*{\frac{S_i}{N_{conv}}}$. Row partitioning is typically only used for the first layer of CNNs with high-resolution inputs. In later layers, the size of inputs usually will be reduced through pooling.

\subsection{Accuracy of row tiling/partitioning} \label{sec:rowtilingaccuracy}
We evaluate the accuracy of the proposed row tiling method with 1D convolution (theoretical accuracy of PhotoFourier) on three common CNNs using ImageNet dataset, which are AlexNet \cite{alexnet}, VGG-16 \cite{vgg16}, and ResNet-18 \cite{resnet}.
Prior works on photonic accelerators that focus on system architectures either did not report any accuracy \cite{shiflett2020pixel, shiflett2021albireo} (accelerate uncompressed neural networks) or just reported theoretical accuracy of their compression method \cite{liu2019holylight, zokaee2020lightbulb} (accelerate compressed neural networks). Therefore the theoretical accuracy is the only metric to compare the relative accuracy between different on-chip photonic accelerators, and we compare them whenever possible in this evaluation. 


\begin{table}[htbp]
  \centering
  \caption{Original accuracy of three CNNs and the accuracy drop of different neural network accelerators (\%). T-1 and T-5 mean top-1 and top-5 accuracy. Ours stands for the proposed row tiling/partitioning method with 1D convolution. Accuracy drop is reported instead of raw accuracy because we have slightly different original accuracy than what is reported in \cite{liu2019holylight} and \cite{zokaee2020lightbulb}. Top-1 accuracy is not reported in both prior works.}
    \begin{tabularx}{\linewidth}{l|XX|XX|X|X}
    \toprule
          & \multicolumn{2}{Y|}{Original}       & \multicolumn{1}{Y}{Ours} &       & \multicolumn{1}{l|}{\cite{liu2019holylight}} & \multicolumn{1}{l}{ \cite{zokaee2020lightbulb}} \\
    \midrule
          & T-1 & T-5 & T-1 & T-5 & T-5 & T-5 \\
    AlexNet & 56.5  & 79.1  & -0.7  & -0.4  & -0.8  & \multicolumn{1}{l}{N/A} \\
    VGG-16 & 73.4  & 91.5  & -0.8  & -0.4  & \multicolumn{1}{l|}{N/A} & \multicolumn{1}{l}{N/A} \\
    ResNet-18 & 69.8  & 89.1  & -1.3  & -0.9  & -0.6  & -1.5 \\
    \bottomrule
    \end{tabularx}%
  \label{tab:accuracyideal}%

\end{table}%
The evaluated accuracy results are shown in Table \ref{tab:accuracyideal}, original accuracy is the floating-point accuracy. We use the row tiling algorithm in this evaluation, but partial row tiling and row partitioning should achieve the same accuracy. In general, PhotoFourier with the proposed row tiling/partitioning method can achieve less than 1\% drop in top-1 and top-5 accuracy for most cases and performs on par with or better than \cite{liu2019holylight} and \cite{zokaee2020lightbulb}. The accuracy results for the row tiling method are inference only using weights trained with 2D convolutions, and the accuracy drop could be eliminated with retraining. 

\section{PhotoFourier Compute Unit} \label{sec:pfcu}
We name the building block of the proposed PhotoFourier accelerator PhotoFourier Compute Unit (PFCU). Each PFCU is essentially an optimized version of the JTC system shown in Section \ref{sec: baselinejtc}. 

\subsection{Pipelining the PFCU}
\begin{figure}[h]
    \centering
    \includegraphics[width=1.0\columnwidth]{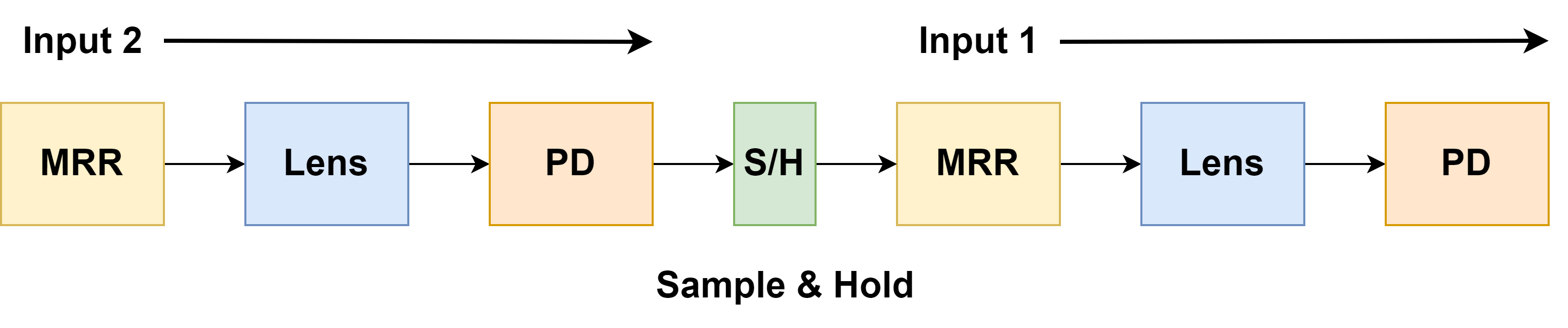}
    \caption{Visualization of pipelined PFCU.}
    \label{fig:jtcpipeline}

\end{figure}

The baseline JTC system requires photodetectors and MRRs in the middle of the system to implement the square function in the Fourier domain, hence the system can be split into two identical parts each with a set of MRRs, Fourier lens, and photodetectors. The reaction time of photodetectors is usually the bottleneck and prevents the system from operating at higher frequencies. Figure \ref{fig:jtcpipeline} depicts the pipelined version of the JTC system. The pipelining is implemented by adding a sample and hold unit at the Fourier plane to buffer the output of the photodetectors. This two-stage pipelined PFCU, processing two convolutions at the same time, can double the throughput with a negligible increase in energy per convolution. 

\subsection{Optimizing PFCU for small filters} \label{sec:pfcuoptimization}
JTC is originally designed to compute the convolution between two input signals of the same size. Therefore, the number of input waveguides is the same as the number of filter waveguides in the baseline JTC system 
However, for CNNs, the filter size is typically much smaller than the size of input activations. 
Each filter waveguide requires a DAC and an MRR to generate the corresponding weight value, and these devices are redundant if the waveguide is inactive, which means the waveguide never needs to generate non-zero values. To improve the area and power efficiency of the JTC system, DACs that correspond to inactive weight waveguides should be removed. Since the locations of active waveguides depend on the input activation size according to the proposed row tiling method, MRRs should not be removed so that every filter waveguide can be active if necessary. MRRs require far less area and power compared to DACs, and can be power gated to save power when inactive. In modern CNNs, the filter size is rarely larger than $5 \times 5$, therefore most of the waveguides are inactive. PFCU keeps 25 active waveguides with corresponding DACs for backward compatibility considerations. For the rare cases where the filter size is larger than $5\times 5$, the inputs and filters can be partitioned to fit onto PFCUs (discussed in Section \ref{sec:partialrowtiling}). The inactive waveguides act as zero-padding, they are still fabricated on the JTC, but they do not receive any inputs and consume zero energy. 

\section{Architecture Design} \label{sec:architecture}
We will introduce the high-level architecture and configuration of PhotoFourier first. The optimizations and reasons behind the choice of design parameters will be covered later in this section (Section \ref{sec:bottleneckanalysis} to Section \ref{sec:dataflowandreuse}).

\subsection{Overall system architecture} \label{sec:overallarchitecture}
\begin{figure*}[h]
    \centering
    \includegraphics[width=1.0\linewidth]{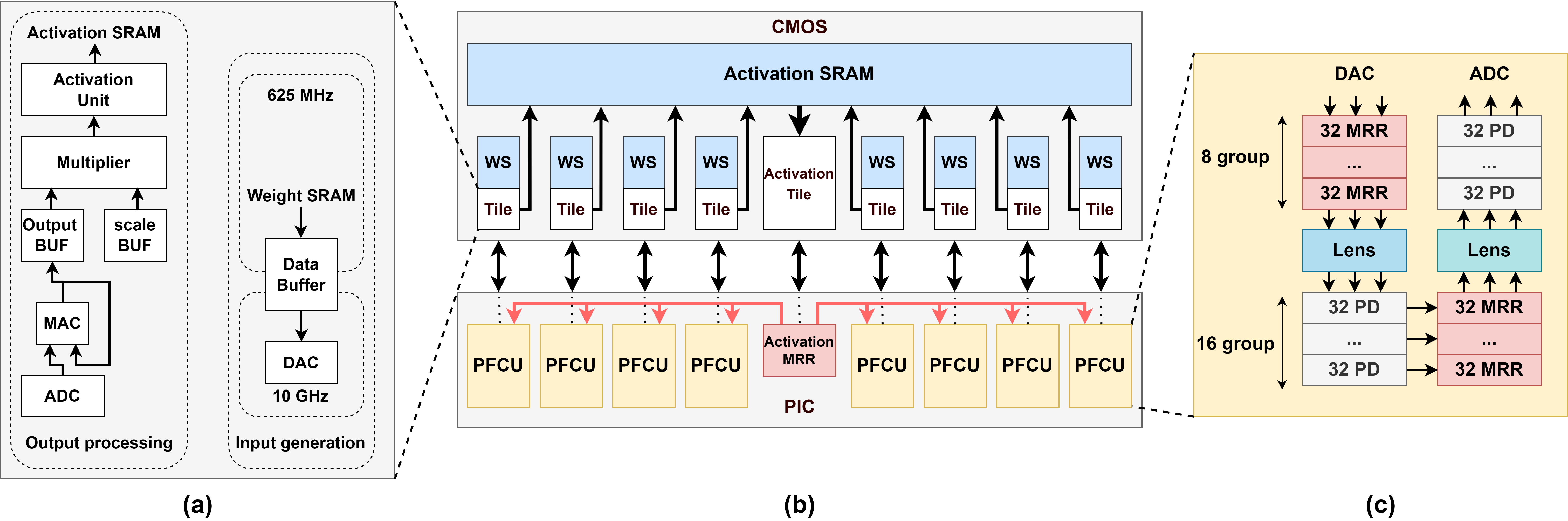}
    \caption{High-level architecture diagram of PhotoFourier-CG. WS stands for weight SRAM, BUF stands for buffer and PD stands for photodetector. (a): CMOS processing tile assigned to one PFCU. It contains two parts, one part for filter weight data generation and the other part for output processing. (b): PhotoFourier-CG architecture with 8 PFCUs, using 2.5D integration. (c): Simplified layout diagram of PFCU, MRRs, and photodetectors are grouped to reduce the width of PFCU.}
    \label{fig:highleveldiagram}

\end{figure*}

We architect two versions of PhotoFourier, PhotoFourier-CG (current generation) and PhotoFourier-NG (next generation). PhotoFourier-CG, as its name suggests, uses conservative estimations on area, power, and integration technology. Figure \ref{fig:highleveldiagram} shows the high-level architecture of PhotoFourier-CG. We architect PhotoFourier-CG as a two-chiplet system, with one CMOS chiplet and one photonic integrated circuits (PIC) chiplet. The PIC contains 8 PFCUs, each with 256 input waveguides, and is clocked at 10 GHz. PhotoFourier-CG by default operates at 8-bit precision.
PhotoFourier-CG uses input broadcasting and OS dataflow, and implements 16-channel temporal accumulation to reduce the ADC and CMOS (except for the input generation circuit) frequency to 625 MHz. The input and weight DACs still operate at 10 GHz while SRAM operates at 625 MHz. Data buffers are used to communicate between two clock domains during input/weight generation. PhotoFourier also contains 8 CMOS tiles that are designed to handle the input and output of PFCUs, as shown in Figure \ref{fig:highleveldiagram} (a). The CMOS tile contains two sub-circuits, one is for input generation and one is for output processing. The input generation circuit has two clock domains, the slower clock is for weight memory access while the faster clock is used to control the DACs. The output processing circuit is used to read and accumulate photodetector outputs as well as to apply scaling/normalization and activation functions. Each CMOS tile has a 512 KB local weight SRAM while the entire PhotoFourier shares a 4 MB global activation SRAM. The activation memory size is set to be large enough to store the activations of common CNNs \cite{vgg16,alexnet,resnet} locally with ping-pong buffering (2 $\times$ maximum activation size), such that the number of DRAM access is minimized and activations storing and loading can happen at the same time. Similarly, the weight SRAM size is set to store the weights of an entire layer of common CNNs \cite{vgg16,alexnet,resnet}. We have taken pseudo-negative processing into account when determining the size of weight SRAM, which will double the storage requirement (see Section \ref{sec:systemsetup}).
PhotoFourier-CG has an activation tile that is similar to the input generation circuit in Figure \ref{fig:highleveldiagram} (a), but is connected to activation SRAM and generates input activations that are shared among all PFCUs. 

We choose to not assume CMOS and photonics can be fabricated on the same chip monolithically with \emph{advanced} technology nodes in PhotoFourier-CG, unlike some other works that are based on such assumption \cite{shiflett2020pixel, shiflett2021albireo, zokaee2020lightbulb, liu2019holylight}. The reason is the current state-of-art commercial available technology for monolithic CMOS and photonics integration can only fabricate 45nm CMOS \cite{gf45nm}, which is several technology nodes behind the state-of-art 5nm technology \cite{tsmc5nm}.

\paragraph{PFCU layout optimization} \label{sec:pfculayout}
Figure \ref{fig:highleveldiagram} (c) shows a simplified layout diagram of the PFCU. Compared to the baseline JTC system in Figure \ref{fig:pichighlevel} (a), the system is flipped after the first set of photodetectors which is in the middle of the system and signals travel towards the CMOS chip in the second part of the system. This folded layout is adopted to place the weight MRRs and the final photodetectors on the same side of the PFCU and close to the CMOS chiplet, which can reduce the length of the analog signals to/from the CMOS chiplet. In the 2-chiplet based system, ADCs and DACs are placed on the CMOS chiplet, hence analog signals need to travel between the chiplets. The loss of analog link due to IR drop will be troublesome if the link length is too long, therefore final photodetectors cannot be placed on the other end of the PIC. 

Another layout optimization is the MRRs and PDs are grouped into rows with the size of 32 and stacked vertically to reduce the PFCU width (if oriented as in Figure \ref{fig:highleveldiagram} (c)). Placing 512 MRRs and PDs in one row horizontally will lead to more than 10 mm width of a single PFCU, which makes the multi-PFCU layout impossible. Even with this optimization, each PFCU still has about 2.32 mm width due to a large number of waveguides and the folded layout. This width makes the layout and fabrication of 16-PFCU challenging as all PFCUs need to be placed close to the CMOS chiplet. Therefore, PhotoFourier-CG only uses 8 PFCUs to make the PIC width reasonable. 

\paragraph{PhotoFourier-NG}
We also architect an advanced version of PhotoFourier, PhotoFourier-NG, which assumes next-generation technologies that are not mature enough currently, but will be available in the near future. On the architecture level, there are two main differences compared to PhotoFourier-CG: (1) PhotoFourier-NG assumes non-linear materials are used to implement the square function of JTC passively instead of photodetectors and MRRs; (2) PhotoFourier-NG assumes monolithic integration of CMOS and photonics with advanced technology node, which eliminates all layout constraints discussed in the previous paragraph. In this case, PFCUs no longer need to have a folded layout and can be placed in just one dimension. Therefore, PhotoFourier-NG uses 16 PFCUs instead of 8 to improve power efficiency. Besides these two differences, all other design parameters are the same for PhotoFourier-CG and PhotoFourier-NG.

\subsection{Bottleneck analysis of baseline system} \label{sec:bottleneckanalysis}
\begin{figure}[h]
    \centering
    \includegraphics[width=0.9\columnwidth]{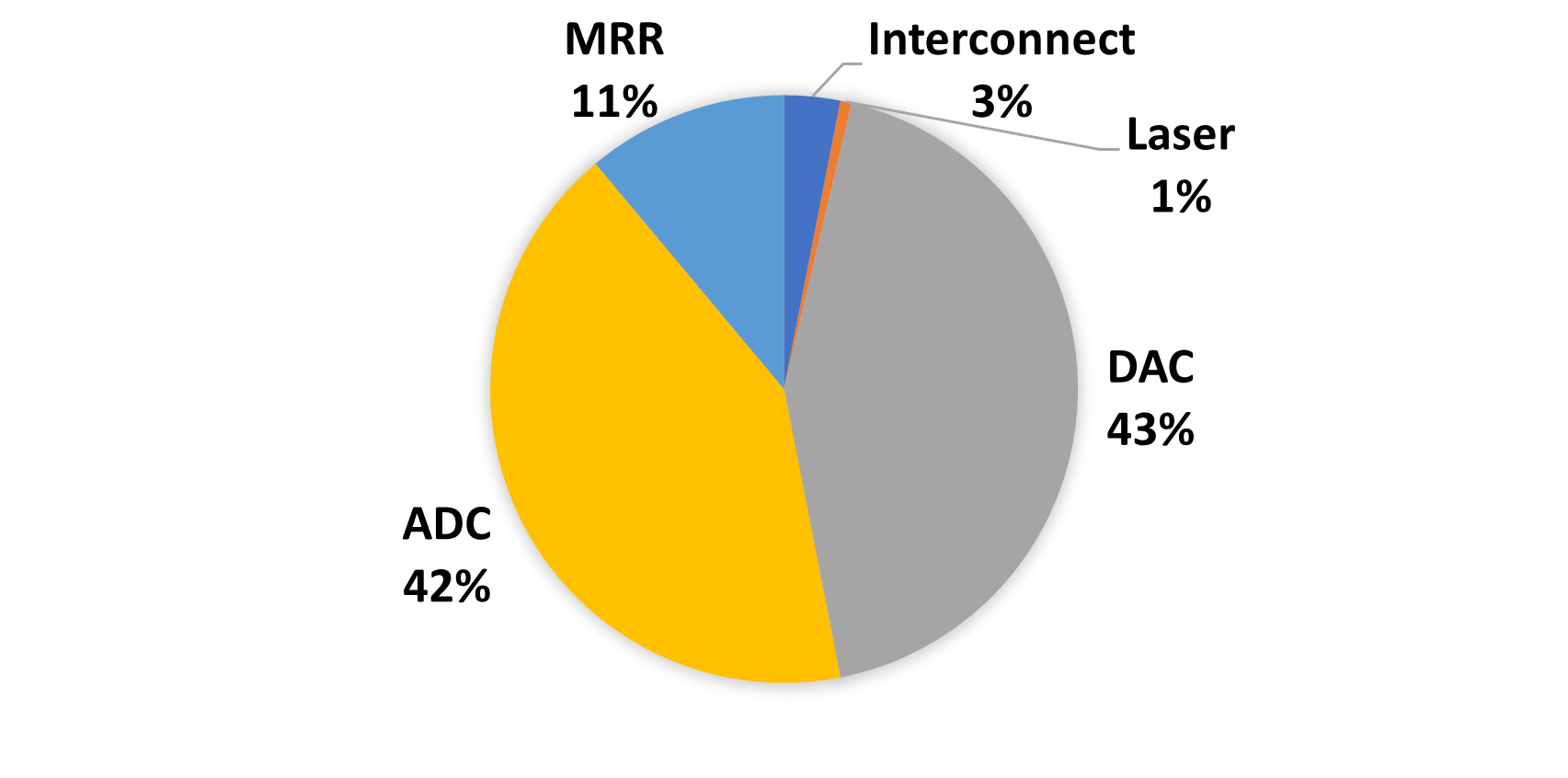}
    \caption{Power contribution of different components of a 1-PFCU baseline system.}
    \label{fig:baselinepowerpiechart}

\end{figure}

To optimize the system for power efficiency, it is important to understand the power bottleneck of a baseline system. The baseline system is configured as having 1 PFCU, 256 input activation waveguides, and clocked at 10 GHz. We evaluate the system on VGG-16 \cite{vgg16} and profile the power contribution of different components. Figure \ref{fig:baselinepowerpiechart} shows the power profiling results. ADCs and DACs dominate the system power and contribute more than 80\% of the total system power. Therefore the architecture and dataflow should be designed to minimize the number of O-E and E-O conversions. 

\subsection{Temporal accumulation} \label{sec:temporalanalysis}
From the results of Section \ref{sec:bottleneckanalysis}, it's clear that O-E and E-O conversions are the bottlenecks of the baseline system. It is crucial to reduce the power consumption of DACs and ADCs to improve overall power efficiency. Thus, DACs and ADCs should be as low-frequency and low-precision as possible (lower frequency also makes the CMOS receiving circuit operate slower). However, lower precision usually leads to worse accuracy, especially for ADCs since they quantize partial sums which typically require higher precision than activations and weights.


To address these issues, we propose the temporal accumulation method, which is a key optimization of PhotoFourier. Temporal accumulation is a method to accumulate the convolution results temporally using photodetectors (before the O-E conversion), which can reduce the ADC and CMOS frequency and the output data bandwidth, as well as improve the accuracy for designs with 8-bit ADCs.
Since the accumulation happens before the ADC readout which applies quantization, temporal accumulation can be considered as full-precision and can improve the accuracy. In other words, temporal accumulation allows 8-bit ADCs to be used without significant accuracy drop, which is otherwise impossible for certain cases. The accumulation happens at the photodetector and can be achieved through capacitors which accumulate the charges to be read at a later time.

In a typical CNN, the convolution results of different input channels need to be accumulated to compute the final output activation, therefore the dataflow needs to be organized in a way that the innermost loop is the input channel such that the output of the channels can be accumulated by the photodetector.
There are both accuracy and performance considerations when choosing the number of channels that are accumulated by the photodetector (temporal accumulation depth). The accuracy results suggest that the temporal accumulation depth of 16 achieves the best accuracy and can restore the accuracy drop due to ADC quantization. Further increasing the temporal accumulation depth will not improve the accuracy, but will make photodetectors larger and slower (and harder to design). Also, for small CNNs, the number of input channels can be quite small and can result in under-utilization if the temporal accumulation depth is too large. Therefore, we set the temporal accumulation depth to 16 in PhotoFourier, which significantly improves overall accuracy, while still being flexible and implementable.
This leads to $16\times$ reduction in ADC frequency, CMOS frequency on the receiving end, and output bandwidth. Consequently, the power of ADC can be massively reduced while the CMOS circuit can operate below 1 GHz (except for the input/weight generation circuit). Temporal accumulation addresses two issues faced by photonic neural network accelerators with minimal hardware overhead, hence is the de-facto design choice of PhotoFourier and is prioritized in our dataflow and parallelization scheme analysis.

\subsubsection{Temporal accumulation accuracy} \label{sec:tempacc}
\begin{figure}[h]
    \centering
    \includegraphics[width=0.95\columnwidth]{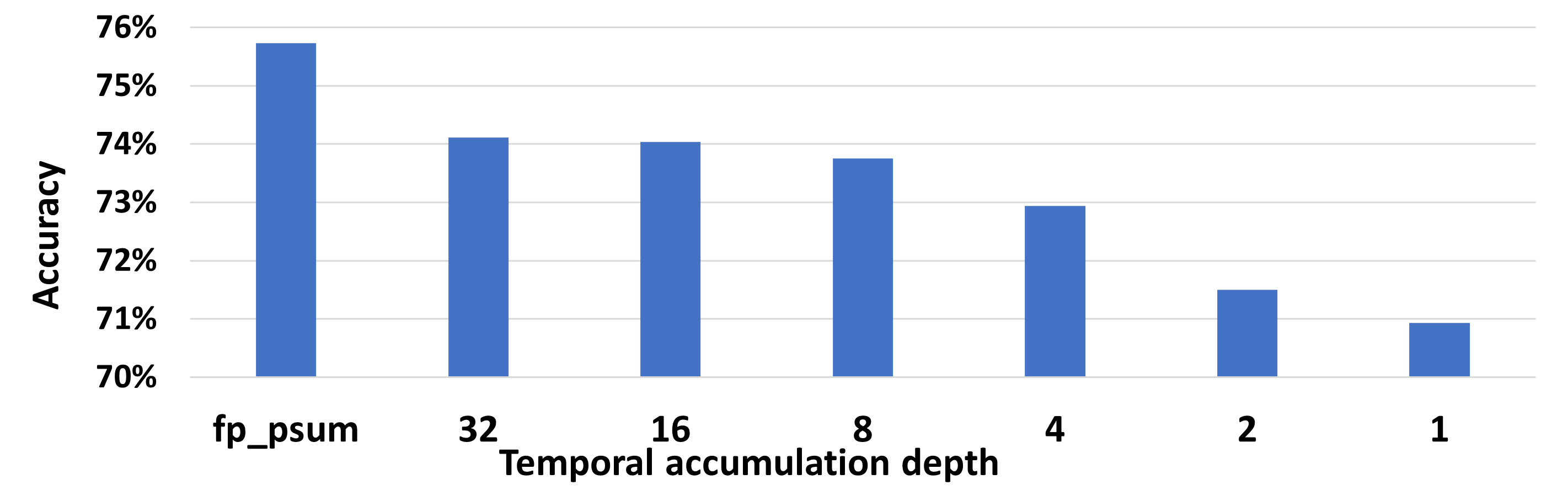}
    \caption{Accuracy of ResNet-s versus different temporal accumulation depth. fp{\_}psum is the accuracy without ADC quantization.}
    \label{fig:tempacc_acc}

\end{figure}

To demonstrate temporal accumulation can improve accuracy, we generate the accuracy results of ResNet-s (a pruned version of ResNet-18 used in \cite{mlperftiny}) on CIFAR-10 with different temporal accumulation depths. ResNet-s is selected since it is a compressed network and is more sensitive to quantization. 
The model simulates the impact of photodetection, which includes applying square function to partial sums and adding sensing noise. The signal-to-noise (SNR) ratio is obtained by computing the average signal power at the photodetectors and compare to the noise power due to dark current. The results in Figure \ref{fig:tempacc_acc} suggest that temporal accumulation can significantly improve the accuracy for designs with 8-bit ADCs. The reason is that 8-bit precision is not enough for partial sums even though it is typically enough for inputs and weights. Since each ADC quantization incurs a large quantization error, having a greater temporal accumulation depth results in fewer partial sum quantization operations (temporal accumulation is full precision), and leads to smaller overall quantization error and better accuracy. 

\subsection{Choice of parallelization scheme} \label{sec:parallelscheme}
We architect PhotoFourier as a system that consists of multiple PFCUs, hence the suitable parallelization scheme needs to be determined for optimal power efficiency. Given a set of available PFCUs, there are three parallelization schemes that can be considered, namely input broadcasting, weight broadcasting, and channel parallelization. Input broadcasting broadcasts the input activations to all PFCUs, and each PFCU computes a unique filter. In this scheme, the DACs and MRRs used to generate input activations can be shared among all PFCUs. Weight broadcasting broadcasts a single filter to all PFCUs, and each PFCU processes a unique convolution window or a full input activation (requires batched processing). Similarly, the DACs and MRRs required to generate filter weights can be shared.In channel parallelization, each PFCU processes one input channel, and the convolution outputs of all PFCUs are accumulated with a single set of photodetectors and ADCs. Channel parallelization scheme shares ADCs among PFCUs rather than DACs and MRRs. 
The three parallelization schemes can also be mix-and-matched for optimal results. 

In our parallelization scheme analysis, we exclude the weight broadcasting scheme and only consider input broadcasting and channel parallelization. There are two reasons for this choice: (1) The number of DACs required to generate filter weights is significantly less than input activations, so the benefit of weight broadcasting is less than input broadcasting; (2) For many situations an entire input activation can be loaded on a single PFCU, thus multi-batch processing is required for weight broadcasting, which is not always possible for inference tasks. Output stationary dataflow is used in this analysis, which is required for temporal accumulation.

\begin{table}[h]
\caption{Table of notations used in the analysis and their meaning.}
\begin{tabularx}{\linewidth}{l|X}
\hline
Notation & Definition                                         \\ \hline
$N_i$    & \# input waveguides of each PFCU         \\
$N_w$    & \#  active weight waveguides of each PFCU \\
$N_{PFCU}$ & \#  PFCUs available                       \\
$IB$     & \#  PFCUs that inputs are broadcasted to  \\
$CP$     & \#  PFCUs that share ADCs                \\
$N_{TA}$     & \# channels accumulated at photodetector             \\
$P_{DAC}$  & Power of DAC                                    \\
$P_{ADC}$  & Power of ADC                          \\ \hline         
\end{tabularx}

\label{tab:notation}

\end{table}

The optimal parallelization scheme can be formulated as a minimization problem of minimizing the sum of ADC and DAC power since they dominate the power consumption. Table \ref{tab:notation} summarizes the notations used in the analysis. Given a fixed $N_{PFCU}$, the design parameter we want to find a solution is $IB$, and $CP$ can be computed by $ CP = \frac{N_{PFCU}}{IB}$. Assuming the power of ADC scales linearly with frequency, the sum of ADC and DAC power can be computed as:
$$P_{total} =P_{ADC} \times \frac{IB\times N_i}{N_{TA}} + P_{DAC} \times (CP\times N_i + N_{PFCU}\times N_w)$$
Since the power of ADC and DAC with the same frequency are similar, they can be removed from the minimization formulation. After some simplification, the minimization problem can be formulated as:
$$ \textbf{Mimimize }  \frac{IB}{N_{TA}} + CP$$
$$ \textbf{Subject to: } IB \times CP = N_{PFCU} $$
This minimization problem can be solved by rewriting $CP$ to $\frac{N_{PFCU}}{IB}$. The exact solution depends on hyperparameters $N_{TA}$ and $N_{PFCU}$. By setting $N_{TA} = 16$, we can sweep $IB$ to find solutions for different $N_{PFCU}$. 
\begin{figure}[h]
    \centering
    \includegraphics[width=0.9\columnwidth]{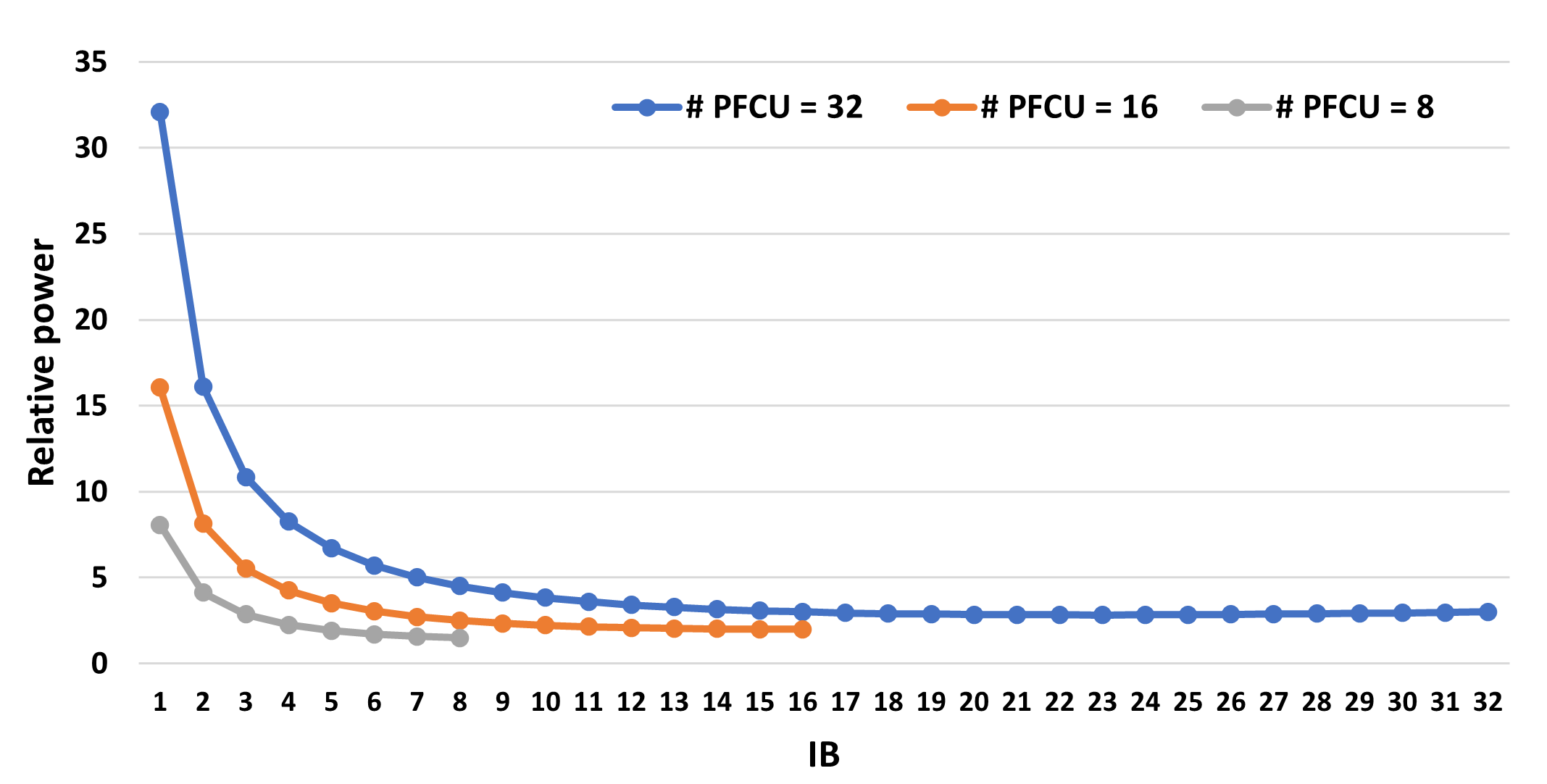}
    \caption{Value of $\frac{IB}{N_{TA}} + CP$ with different $IB$, for three different number of PFCUs.}
    \label{fig:IBsweep}
\end{figure}

Figure \ref{fig:IBsweep} plots the value of $\frac{IB}{N_{TA}} + CP$ for all $IB$ values. For the cases where $N_{PFCU}$ is 8 or 16, the system power minimizes when $IB = N_{PFCU}$. When $N_{PFCU} = 32$, the system power is same when $IB = 16$ and $IB = 32$, and the minimum system power is achieved when $IB = 23$. However, the solution is not valid as the valid solutions for $IB$ can only be $1,2,4,8,16,32$ (due to the constraints of $IB$ and $CP$). Therefore, both 16 and 32 are optimal solutions of $IB$ when $N_{PFCU} = 32$. The result suggests that when $N_{TA}$ is set to 16, and the number of PFCUs is less than or equal to 32, input broadcasting is always the best parallelization scheme.

\subsection{Number of waveguides and PFCUs} \label{sec:numwaveguidebenchmark}
The number of input waveguides per PFCU and the total number of PFCUs are two design parameters that need to be determined, and they are discussed together because of the trade-off between them. Given a fixed area budget, the more waveguides per PFCU, the fewer number of PFCUs can be placed. Assuming input broadcasting is used, increasing the number of PFCUs improves the power efficiency by sharing the input activation with more filters, while increasing the number of waveguides per PFCU can also improve the power efficiency by effectively sharing a filter with more convolution windows (weight broadcasting within PFCU). However, PFCU can be under-utilized if the number of input waveguides is too large. Under-utilization typically happens when the system is executing later layers of a CNN, where the input activation size can be small (caused by pooling). For example, assuming the number of input waveguides is 512, then the PFCU will not be fully utilized if the input size is less than $23\times 23$, which is common in the later layers of CNNs (e.g., ResNet-34 has 18 convolution layers with input size $\leq 14\times 14$). 

The optimal choice of the number of waveguides and the number of PFCUs is determined through benchmarks on real CNNs. We set the area budget of PhotoFourier-CG to 100 $mm^2$, which is around the upper limit of chip area due to the layout constraint discussed in Section \ref{sec:pfculayout}. For consistency, we use the same area budget for PhotoFouirer-NG.
We select 5 values for the number of PFCU and compute the maximum number of input waveguides per PFCU under the area budget for both PhotoFouirer versions. We then evaluate the selected configurations on 5 different CNNs, which are AlexNet \cite{alexnet}, VGG-16 \cite{vgg16}, ResNet-18 \cite{resnet}, ResNet-32, and ResNet-50, and compute the geometric mean of the normalized FPS/W on these 5 CNNs. The results are listed in Table \ref{tab:n_waveguidevsFPCU}. PhotoFourier-CG achieves the best average FPS/W with 8 PFCUs and 270 waveguides per PFCU, while PhotoFourier-NG achieves the best average FPS/W with 16 PFCUs and 267 waveguides per PFCU. Therefore, for the given area budget, we architect PhotoFourier-CG to have 8 PFCUs with 256 input waveguides per PFCU and PhotoFouirer-NG to have 16 PFCUs with 256 input waveguides per PFCU. 



\begin{table}[h]
\caption{Maximum number of input waveguides per PFCU and the geometric mean of normalized FPS/W on 5 CNNs for PhotoFourier-CG and PhotoFourier-NG with different number of PFCUs, given a 100 $mm^2$ area budget.}
    \begin{tabularx}{\linewidth}{r|cc|cc}
    \toprule
          & \multicolumn{2}{c|}{PhotoFourier-CG}      & \multicolumn{2}{c}{PhotoFourier-NG} \\
    \midrule
    \multicolumn{1}{l|}{\# PFCU} & \multicolumn{1}{l}{\# waveguides } & \multicolumn{1}{l|}{avg. FPS/W} & \multicolumn{1}{l}{\# waveguides } & \multicolumn{1}{l}{avg. FPS/W} \\
    \midrule
    4     & 412   & 0.70  & 576   & 0.55 \\
    8     & 270   & \textbf{0.97}  & 395   & 0.75 \\
    16    & 172   & 0.89  & 267   & \textbf{0.97} \\
    32    & 105   & 0.72  & 177   & 0.82 \\
    64    & 61    & 0.74  & 114   & 0.81 \\
    \bottomrule
    \end{tabularx}%
\label{tab:n_waveguidevsFPCU}

\end{table}

\subsection{Dataflow and reuse} \label{sec:dataflowandreuse}
\subsubsection{Reuse analysis}
Output stationery (OS) dataflow is used in PhotoFourier to implement temporal accumulation. In OS dataflow, every cycle processes a new channel of input activations and filters, so that the convolution results (partial sums) can be accumulated locally at the photodetectors. OS dataflow minimizes output data bandwidth at the cost of higher input/weight bandwidth.
If input broadcasting is used without any data reuse scheme (e.g., OS dataflow), then the output bandwidth will be $8$-$16\times$ higher than the input bandwidth, which may prevent the architecture to scale efficiently. OS dataflow addresses the imbalance between input and output bandwidth. By accumulating 16 channels at the photodetector, output bandwidth reduces by $16\times$. Although the input broadcasting $+$ OS dataflow scheme does not lead to a direct reduction in weight bandwidth, the weight bandwidth requirement is much smaller than the input and output bandwidth for a sigle PFCU. There are two factors that contribute to the small weight bandwidth. (1): filters are much smaller than activations in general. (2): the local weight reuse over different convolution windows within each PFCU. This reuse is inherently implemented by JTC, which computes an entire convolution in one cycle so the weights are effectively shared among the inputs loaded onto the JTC. The weight reuse can be seen as weight broadcasting within the PFCU. When executing a $3\times 3$ convolution layer, the total weight bandwidth can be $1.78\times$ lower than the input and output bandwidth. Overall, PhotoFourier utilizes data reuse (sharing) on all three dimensions, which are input, weight, and partial sum. Input reuse is achieved by broadcasting the inputs to all PFCUs, weight reuse is achieved inherently by the JTC, and partial sum reuse is achieved by temporal accumulation at photodetectors. 

\subsubsection{Execution sequence}
During the execution, PhotoFourier processes 8/16 filters in parallel and computes the convolution of one input channel tile per cycle. After completion of one cycle, PhotoFourier processes the next input channel, until all input channels are processed. Since many convolution layers have more than 16 input channels, two-level accumulation is implemented. Input channels are divided into groups of 16 and partial sums are accumulated by the photodetector within each group, while CMOS accumulators are used to accumulate partial sums between groups. Once all input channels are processed, the accumulated results are sent to activation units to generate the output activation, which will then be stored in the activation memory. This process will be repeated until all the filters and input tiles are processed.

\begin{figure}[htbp]
    \centering
    \includegraphics[width=0.9\columnwidth]{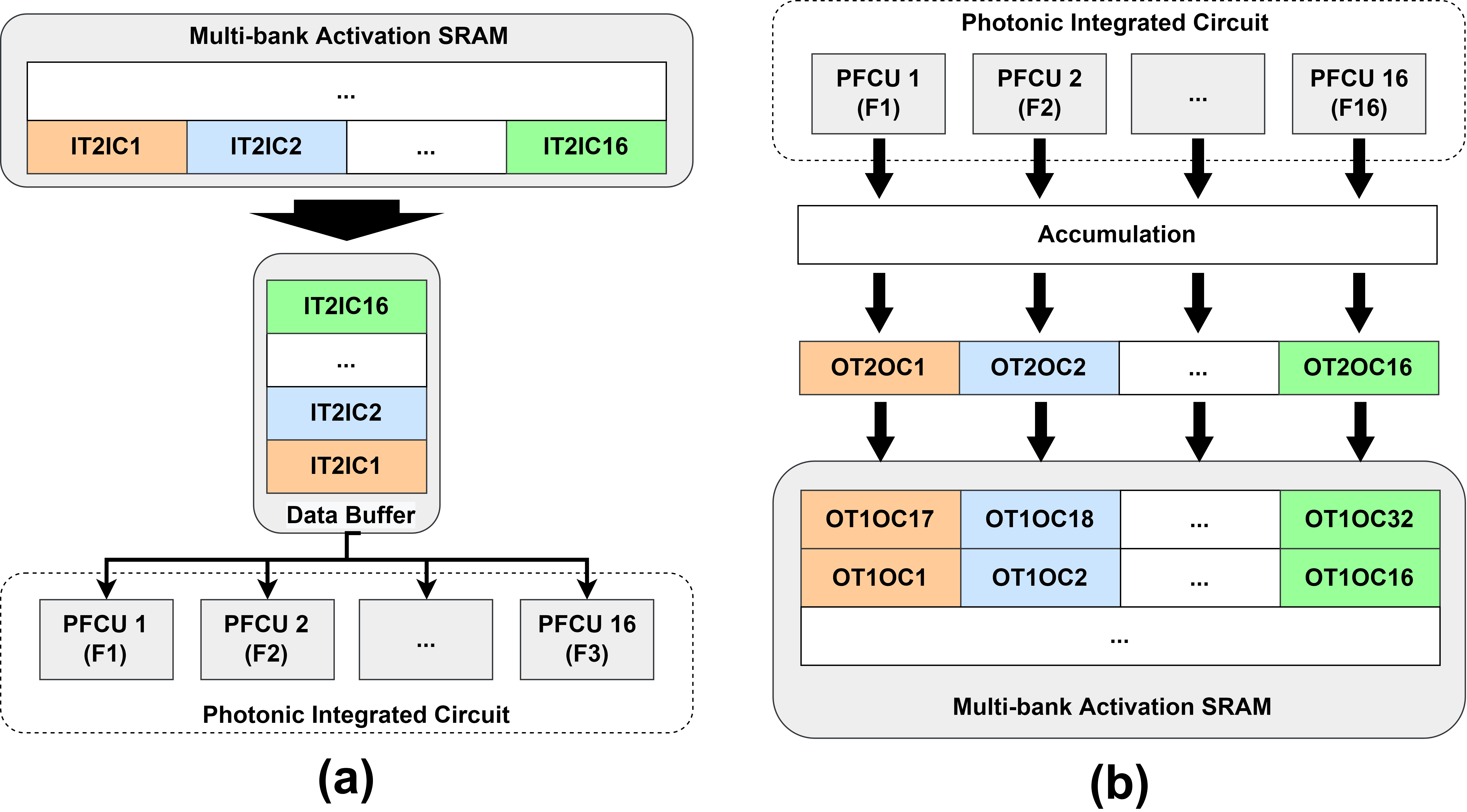}
    \caption{Visualization of the activation memory mapping and the execution process of input tile 2. IT: input tile, OT: output tile, IC: input channel, OC: output channel, F: filter. (a): Accessing input activation from memory. (b): Storing output activation to memory.  }
    \label{fig:memorder}

\end{figure}

Figure \ref{fig:memorder} visualizes the activation memory mapping and the execution process of one input tile (tile 2). 16 channels of the input tile are loaded from the activation memory to the data buffer in parallel. This is possible since they are contiguous in memory. The data buffer connects the slower clock domain to the faster clock domain. The data buffer loads the 16 channels in one cycle of the slow clock domain and the input generation circuit operates 16X faster to generate one channel of the input tile in every cycle of the fast clock domain, which is then broadcasted to all PFCUs. Each PFCU processes a unique filter so 16 filters are processed in parallel, which produces the partial sums of 16 output channels. 

\section{Evaluation}
\subsection{System Setup} \label{sec:systemsetup}

We build a custom Python-based simulator to simulate the latency, power, area, and efficiency of PhotoFourier on actual CNN inferences. For PhotoFourier-CG, we use Cadence Genus with a commercial 14nm library to simulate the delay, power, and area of the CMOS circuit of PhotoFourier shown in Figure \ref{fig:highleveldiagram} (a). We use a commercial 14nm memory compiler to obtain the area, leakage power, and access energy of activation and weight SRAMs. For PhotoFourier-NG, we scale the technology node to 7nm, using the scaling equations proposed in \cite{nodescale} (for CMOS circuit, based on our 14nm results). We use PCACTI \cite{fincacti} to model the 7nm FinFET SRAM.
The simulation results of the CMOS circuit are embedded into the python simulator and combined with the simulation results of photonics to obtain the performance of the entire accelerator. 
Table \ref{tab:powernumber} lists the power of different components that we used in our simulator and the high-level design parameters, for both PhotoFourier-CG and PhotoFourier-NG. Since there are no references of ADC and DAC with the exact bitwidth, frequency, and technology node, we find 8-bit ADC and DAC with similar technology nodes (14nm and 16nm), but with higher frequency than required. We then linearly scale down the power of the cited ADC/DAC according to the frequency ratio to obtain the value we used in PhotoFourier-CG.
We scale the power of ADC by $5.81\times$ in PhotoFourier-NG, which is obtained using the Walden ADC Figure-of-merit (FOM) formula \cite{waldenfom} with the envelope line (best achievable FOM of published ADCs for different frequencies) constructed in \cite{adcsurvey}. We obtain the best FOM for 625 MHZ to estimate the optimal 8-bit ADC power from the FOM equation. We scale the DAC with the same number since DAC and ADC have similar scaling properties (SAR-ADCs are based on DACs).
Table \ref{tab:areanumber} lists the dimension of the optical components we used to compute the PFCU area, and we keep them the same for both PhotoFourier-CG and PhotoFourier-NG. The laser power is set to maintain larger than 20 dB SNR at photodetectors in most cases, estimated using the dark current of photodetectors and the system loss of PhotoFouirer.
The simulator implements the proposed row tiling/partitioning algorithm when simulating CNN inferences, and uses a batch size of 1. As PhotoFourier is designed as a convolution accelerator, only convolution layers are accelerated and benchmarked. This will not affect the overall speedup on common CNNs\cite{vgg16, resnet} since more than $99\%$ of total MAC operations are from convolution layers.
We use PyTorch with custom convolution functions to generate all the accuracy results used in this paper.


\begin{table}[htbp]
\footnotesize
  \centering
  \caption{Power of different components and high-level design parameters used for PhotoFourier-CG and PhotoFourier-NG}
    \begin{tabularx}{\linewidth}{l|YY}
    \toprule
    \multicolumn{1}{c}{} & PhotoFourier-CG & PhotoFourier-NG \\
    \midrule
    \midrule
    \multicolumn{3}{c}{Component power}  \\
    \midrule
    MRR   & 3.1mW\cite{albireomrr} & 0.42mW\cite{advancedmrr} \\
    \midrule
    Laser  & 0.5mW per waveguide & 0.5mW per waveguide \\
    \midrule
    ADC $@$ 625 MHz& 0.93mW\cite{14nmadc_new}  & 0.16mW \\
    \midrule
    DAC $@$ 10 GHz & 35.71mW \cite{16nmdac_new}  & 6.15mW \\
    \midrule
    \multicolumn{3}{c}{High-level design parameters}  \\
    \midrule
    \# PFCUs & 8     & 16 \\
    \midrule
    \# input waveguides & 256   & 256 \\
    \midrule
    \# chiplets & 2     & 1 \\
    \midrule
     technology node & 14nm & 7nm \\
    \bottomrule
    \end{tabularx}%
  \label{tab:powernumber}%
\end{table}%

\begin{table}[htb]
\footnotesize
  \centering
  \caption{Dimensions of the photonic components used in area estimation}
    \begin{tabularx}{\linewidth}{YY}
    \toprule
    Component & Dimension \\
    \midrule
    MRR\cite{aimphotonics}   & 15 um $\times$ 17 um \\
    Optical splitter\cite{albireoyjunction} & 1.2 um $\times$ 2.2 um \\
    Photodetector\cite{aimphotonics} & 16 um $\times$ 120 um \\
    Waveguide pitch\cite{waveguidepitch} & 1.3 um \\
    Laser \cite{laseralb} & 400 um $\times$ 300 um \\
    On-chip lens & 2 mm $\times$ 1 mm \\
    \bottomrule
    \end{tabularx}%
  \label{tab:areanumber}%

\end{table}%

\emph{Dealing with negative weights:} PhotoFourier uses the pseudo-negative method \cite{standford4f} to deal with negative weights, which can be troublesome for photonic accelerators to process. The pseudo-negative method breaks every filter into a pair of positive-value filters using the formula $x = p - n$, where $x$ is the original weight and $p,n$ are two positive-value filters. Pairs of filters are processed as normal in PFCUs and they are subtracted digitally in the CMOS circuit. The method makes photonic accelerators able to process negative weights but at the cost of $2\times$ computation.

\subsection{Effect of optimizations}

\begin{figure}[htbp]
    \centering
    \includegraphics[width=0.9\columnwidth]{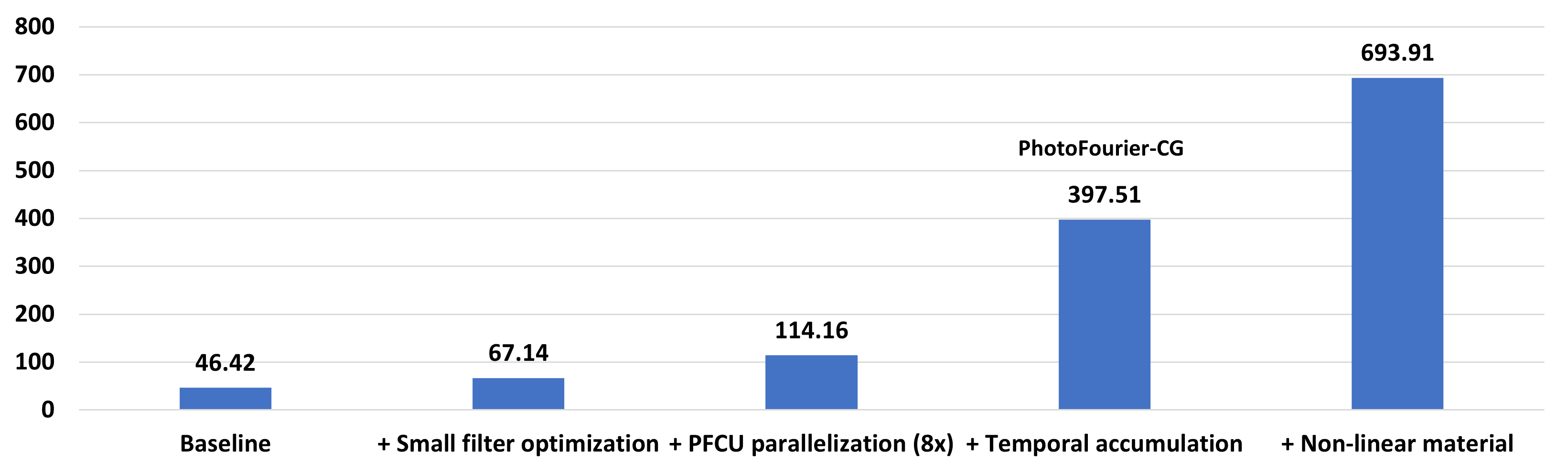}
    \caption{Geometric mean of FPS/W for PhotoFourier with different optimizations. Starting from the baseline, each column adds one optimization to the system, and includes all previous optimizations (columns on the left).}
    \label{fig:opttrend}

\end{figure}

We first demonstrate the effect of proposed optimizations in terms of the geometric mean of FPS/W on the same five CNNs used in Table \ref{tab:n_waveguidevsFPCU}, and the results are shown in Figure \ref{fig:opttrend}. The baseline system is a single-PFCU system with 256 input channels, and we stick with the power number of PhotoFourier-CG in this evaluation to exclude the effect of technology scaling. We order the optimizations from PFCU-level optimization to architectural-level optimization.
Small filter optimizations reduce the number of weight DACs per PFCU, PFCU parallelization shares 256 input DACs with 8 PFCUs, temporal accumulation reduces ADC frequency by $16\times$, and non-linear material (used in NG version only) removes the MRRs used to compute the square function.
The proposed optimizations significantly improve the power efficiency, and can be $15\times$ better than the baseline system.

\subsection{Area}

\begin{figure}[htbp]
    \centering
    \includegraphics[width=0.9\columnwidth]{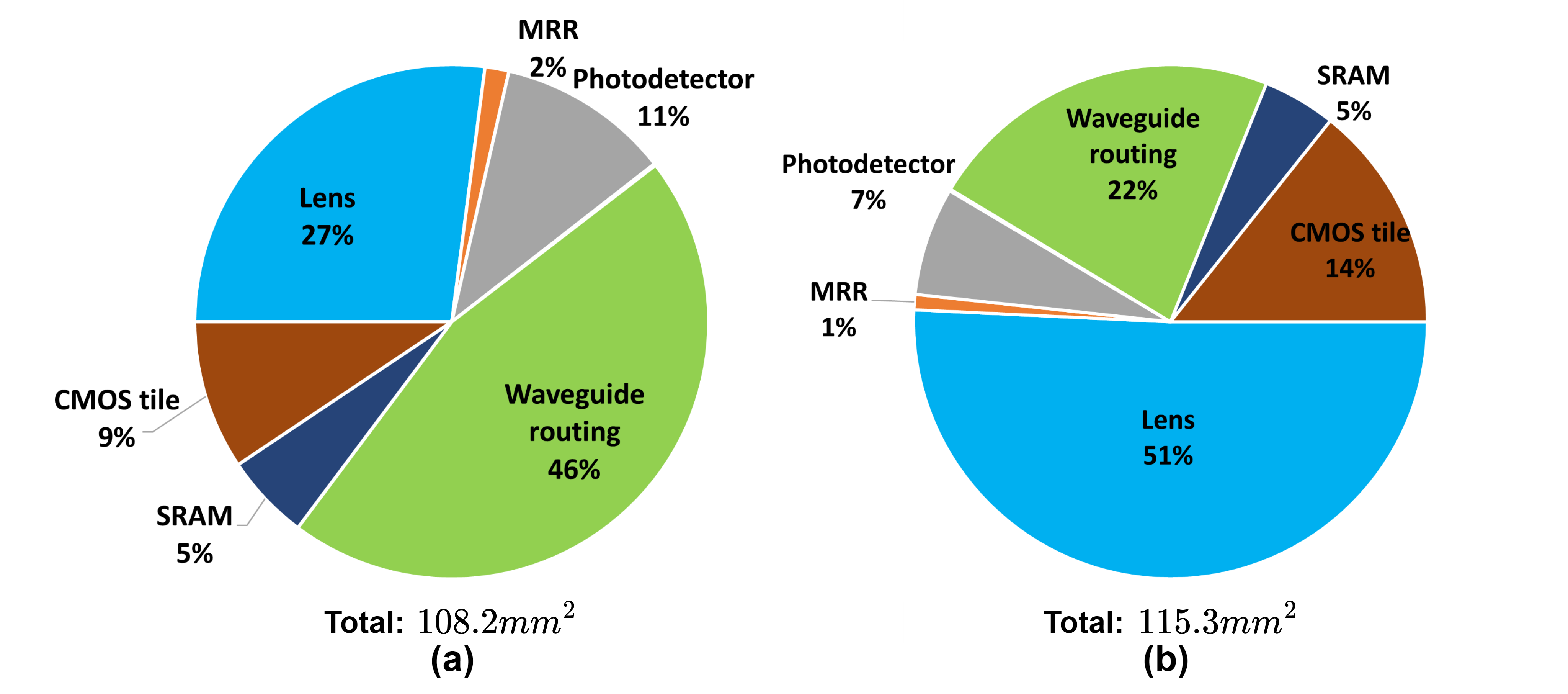}
    \caption{Area breakdown of PhotoFourier. Waveguide routing includes waveguides area and redundant area due to layout constraints. (a): CG version. PIC chiplet: $92.2 mm^2$, SRAM: $5.85 mm^2$, CMOS tile: $10.15 mm^2$. (b): NG version. PFCU: $93.5 mm^2$, SRAM: $5.3 mm^2$, CMOS tile: $16.5 mm^2$.}
    \label{fig:areabreakdown}

\end{figure}

Figure \ref{fig:areabreakdown} shows the total area and area breakdown for two PhotoFourier versions, with photonic components dominating the area for both versions. While having $2\times$ PFCUs, PhotoFourier-NG has roughly the same area as PhotoFourier-CG. The area reduction of individual PFCUs is the result of using non-linear materials to replace MRRs and photodetectors that implement the non-linear function. 
For PhotoFourier-CG, waveguide routing (including redundant space) uses nearly half of the chip area. Such low area utilization is caused by the layout constraint discussed in Section \ref{sec:pfculayout}, which makes the layout less compact. In PhotoFourier-NG, the monolithic integration of CMOS and photonics, together with non-linear materials, greatly relaxes the layout constraint and makes the layout more compact. Since photodetector and MRR  consume a very small portion of the total area in both versions, shrinking their sizes can barely improve area efficiency. Instead, more compact on-chip lenses should be studied to further reduce the footprint.

\subsection{Power}

\begin{figure}[htbp]
    \centering
    \includegraphics[width=0.9\columnwidth]{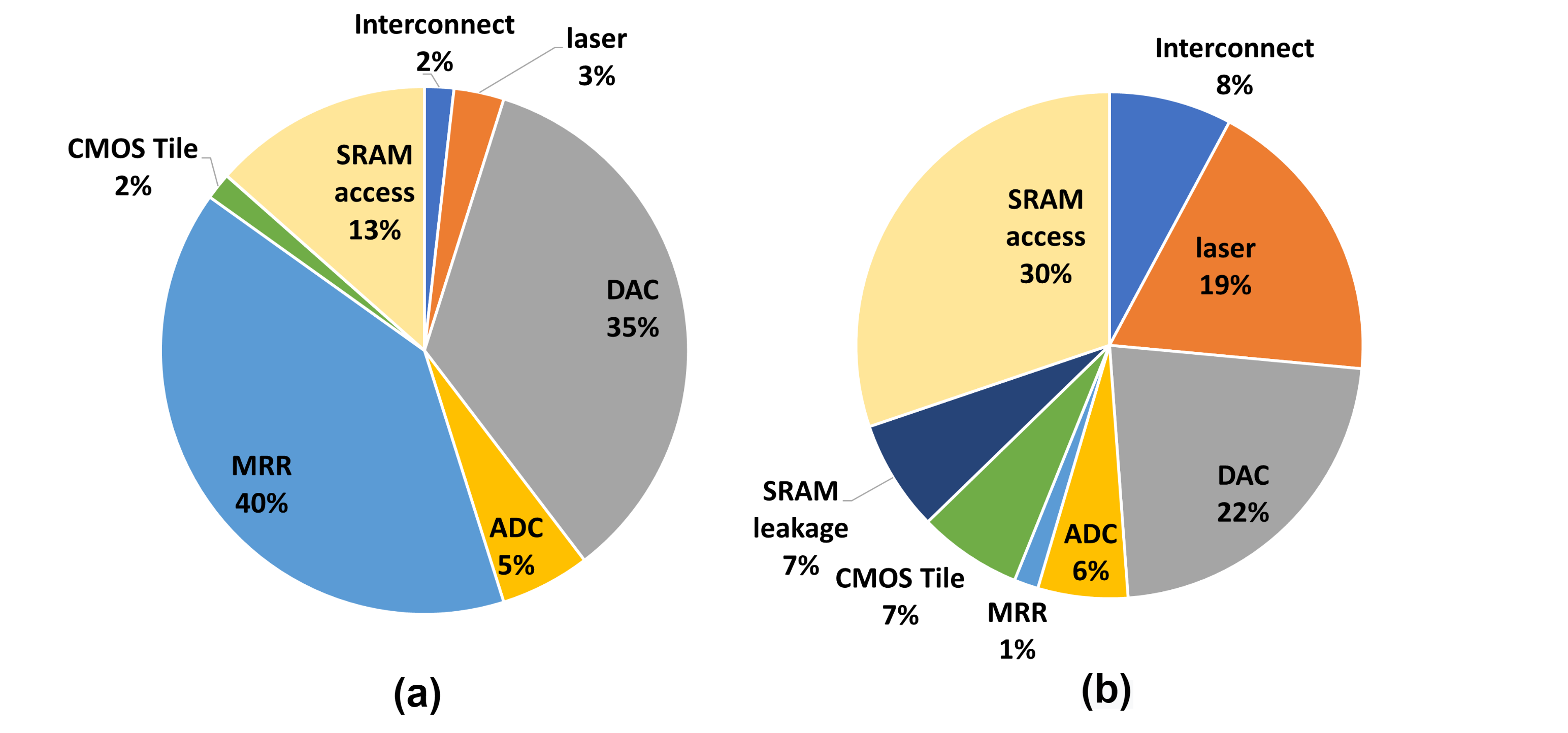}
    \caption{Power breakdown of the two PhotoFourier versions. (a): PhotoFourier-CG. (b): PhotoFourier-NG.}
    \label{fig:powerpiechartcombined}
  
\end{figure}

We benchmark the two versions of PhotoFourier on the same 5 CNNs used in Section \ref{sec:numwaveguidebenchmark} to evaluate their performance. The average power of PhotoFourier CG and NG over the five networks are 26.0 W and 8.42 W respectively. Figure \ref{fig:powerpiechartcombined} shows the power breakdown of PhotoFourier-CG and PhotoFourier-NG. The power distribution of PhotoFourier-CG is somewhat evenly spread across MRR, DAC, and other components. The DAC and ADC no longer dominate the power consumption when compared to a baseline JTC system (Figure \ref{fig:baselinepowerpiechart}), as temporal accumulation and input broadcasting greatly reduce the power of ADCs and DACs. Temporal accumulation can reduce ADC power by more than $30\times$ compared to 10 GHz ADCs \cite{10gsadc}, which makes ADC power significantly less than DAC power.
For PhotoFourier-NG, the SRAM access power replaces MRR/DAC to become the largest contributor to the total system power. There are two reasons, one is the power of MRRs, DACs, and ADCs is further reduced in the NG version, and the reduction is larger than SRAM power reduction due to technology node scaling. Another reason is the SRAM access energy for PhotoFourier is on the higher end, as wide memory buses are required to keep up the bandwidth requirement of the 10 GHz photonic circuits, which increases the access energy. 

We make the following observations: Since data movement (memory accessing $+$ interconnect) dominates the power of PhotoFourier-NG, the priority of further improving power efficiency with next-generation technologies is no longer optimizing O/E conversions. Reducing the data movement cost should be the main focus, which we will discuss more in Section \ref{sec:discussion}.


\begin{figure*}[htbp]
    \centering
    \includegraphics[width=1.0\linewidth]{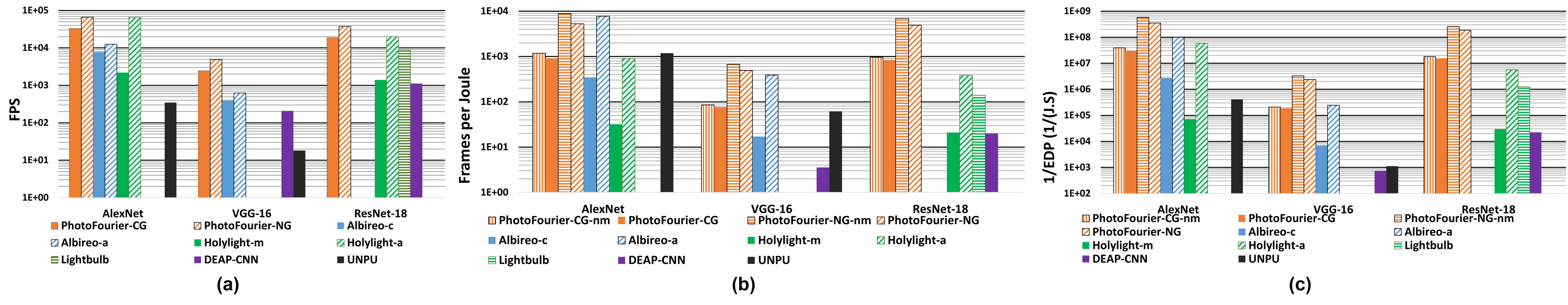}
    \caption{Inference performance of common CNNs on ImageNet dataset, compared to prior works. Missing bars indicate the data are not available from the original papers. (a) Inference throughput in terms of frames per second (FPS). (b): Inference efficiency in terms of frames per Joule (FPS/W). -nm means PhotoFourier versions without memory access power. To the best of our knowledge, Albireo does not report memory access power, so -nm versions are included for reference. (c): Energy-delay Product (EDP) in terms of $\frac{1}{EDP}$, inverse is used for better visualization, therefore larger is better.}
 
    \label{fig:barchartcombined}
    
\end{figure*}

\subsection{Comparison with prior works}

We mainly compare the performance of PhotoFourier with the recently published Albireo accelerator which reports state-of-art power efficiency results for uncompressed CNNs. For reference purpose, we also compare with some other photonic neural network accelerators (DEAP-CNN \cite{DEAPCNN2019}, Lightbulb \cite{zokaee2020lightbulb}, two versions of Holylight \cite{liu2019holylight}) and one digital accelerator (UNPU \cite{UNPU}). 
Albireo is a CNN accelerator based on MZIs and MRRs. DEAP-CNN, Lightbulb, and Holylight are based on only microdisks/MRRs. Albireo and Holylight-m target 8-bit CNNs, DEAP-CNN targets 7-bit CNNs, while Holylight-a and Lightbulb target power-of-two quantized and binary CNNs respectively. Therefore, PhotoFourier is best compared to Albireo and Holylight-m since they can accelerate uncompressed 8-bit CNNs, which is not possible for Holylight-a and Lightbulb.  
We benchmark PhotoFourier on AlexNet, VGG-16, and ResNet-18 as each of the selected accelerators reports results on some of these networks. We compare with both Albireo-c (conservative) and Albireo-a (aggressive), which is similar to PhotoFourier-CG and PhotoFourier-NG. Albireo has more aggressive assumptions for their advanced version since Albireo-a reduces the ADC and DAC power by $10\times$ compared to Albireo-c, while we reduce them by $5.8\times$ in the NG version based on FOM analysis. All results of other works are obtained directly from the original papers without modification except for DEAP-CNN (targets for small CNNs), for which we use a scaled version to generate results for our benchmarks. The results of Holylight and Lightbulb are estimated based on bar charts and are not precise. Albireo uses a 7nm library to model the CMOS components and UNPU uses 65nm technology node. 

Figure \ref{fig:barchartcombined} (a) shows the throughput results in terms of frames per second (FPS). PhotoFourier-CG and PhotoFourier-NG have $5$-$10\times$ higher throughput compared to Albireo-c and Albireo-a. Given that Albireo's chip area ($124.6 mm^2$) is similar to PhotoFourier, PhotoFourier has $5$-$10\times$ better area efficiency than Albireo. Holylight-a and Lightbulb have higher throughput in general since they target quantized CNNs, but are still less than PhotoFourier-NG, except for AlexNet where PhotoFourier-NG is on par with Holylight-a. PhotoFourier is less efficient on AlexNet due to its first convolution layer with $11\times 11$ filter and stride of 4. PhotoFourier is less efficient when processing strided convolutions as PhotoFourier handles them by computing with unit stride and then discarding unnecessary results, limited by the underlying JTC operation which only supports unit stride convolution.

Figure \ref{fig:barchartcombined} (b) shows the FPS/W result, which measures the power efficiency. 
Compared to 8-bit accelerators and with \emph{memory access power modeled}, PhotoFourier-CG achieves around $3$-$5\times$ higher FPS/W than Albireo-c on benchmarked networks, and is $532\times$ and $704\times$ better than Holylight-m and DEAP-CNN respectively. Compared to Albireo-a, PhotoFourier-NG is slightly ahead for VGG-16, but is slightly behind for AlexNet because of the inefficiency of strided convolutions. PhotoFourier-NG achieves similar power efficiency compared to Albireo-a, despite modeling memory access power and using less aggressive ADC/DAC scaling.
Even when compared to Holylight-a and Lightbulb which target heavily quantized CNNs, both PhotoFourier versions achieve better FPS/W, demonstrating superior power efficiency. While having low throughput, UNPU achieves decent power efficiency and is on par with PhotoFourier-CG (but behind PhotoFourier-NG).
Figure \ref{fig:barchartcombined} (c) visualizes the energy-delay product (EDP) in terms of $\frac{1}{EDP}$. Given PhotoFourier's high throughput and power efficiency, PhotoFourier-NG achieves the best EDP on all three networks. Even PhotoFourier-CG has better EDP than other accelerators in most cases, except for the less efficient AlexNet where it falls behind Holylight-a, which targets heavily quantized networks.

We also compare PhotoFourier with another recent MRR-based photonic NN accelerator CrossLight \cite{crosslight}. Since CrossLight does not evaluate using our selected networks, we evaluate PhotoFourier on their custom 4-layer CNN for the CIFAR-10 dataset. PhotoFourier-CG achieves more than $100 \times$ better energy per inference ($4.76 \mu$J vs $427 \mu$J), despite having relatively low utilization on this network. 
Overall, PhotoFourier achieves state-of-art throughput and efficiency results of photonic neural network accelerators. Compared to Albireo which also targets uncompressed CNNs, PhotoFourier-CG achieves up to $28\times$ better EDP compared to Albireo-c and PhotoFouier-NG achieves up to $10\times$ better EDP compared to Albireo-a. The better performance is contributed by the complexity reduction of Fourier optics, as well as the proposed optimizations. PhotoFourier requires fewer optical components to perform the same convolution operation which reduces the area and power of photonic components. Being compact means PhotoFourier can have more components than Albireo with a similar area budget, hence can benefit more from parallelism.

\section{Discussion} \label{sec:discussion}
We discuss some technologies and challenges that are not implemented or adequately addressed in PhotoFourier, to provide potential directions for future works that can make PhotoFourier and photonic NN accelerators in general more efficient.


\textbf{Data movement:} Although PhotoFourier leverages various reuse to reduce the amount of data movement, it still consumes more than 30\% of total system power in PhotoFourier-NG (Figure \ref{fig:powerpiechartcombined}). On one hand, this suggests that the NG version makes compute so efficient that memory access power becomes the power bottleneck of compute-bound CNN inferences. On the other hand, it shows the importance of further reducing data movement cost to continue scaling JTC-based (and other) photonic accelerators. There are both hardware and software approaches to achieve that. On the hardware level, photonic memory \cite{photonicmemory14, photonicmemory21, photonicmemorynature} and interconnects \cite{photonicinterconnect1, photonicinterconnect2, photonicinterconnect3} should be further studied since photonics has no RC delay or $I^2R$ loss. Besides, 3D integration can also reduce the data movement cost.
On the software level, compression methods like quantization, pruning, and parameter sharing can be used to reduce the total number of bits that needs to be transferred.

\textbf{3D integration:} A potential way to further improve the area and power efficiency of PhotoFourier is 3D integration, which allows memory to be stacked on top of compute units to reduce the data movement cost by making interconnects shorter. Thermal management is one of the main challenges of 3D integration \cite{3dicissue1,3dicissue2} because of the increased transistor density. However, photonics generate less heat than CMOS, which makes them more suitable for 3D integration and worth further exploring.

\section{Related work} \label{sec:relatedwork}
4F systems require spatial filters to be transformed into complex-valued Fourier filters before feeding into the system. This requires 4F systems to support complex multiplication, which is hard to implement as it requires both amplitude and phase modulation. Furthermore, 4F systems require filter sizes to match input activation sizes (for point-wise multiplication in the Fourier domain), thereby wasting substantial weight modulation bandwidth (conventional CNNs all have small filters). Unlike 4F systems, JTC treats filters the same way as inputs, where the Fourier lens computes the Fourier transform of filters. Thus JTC can use real-valued spatial filters with arbitrary size (with zero padding) and significantly improves overall efficiency compared to 4F systems.
4F CNN accelerators \cite{standford4f, miscuglio2020optica, 4farchitecurspie} are most closely related to PhotoFourier. However, they all target free-space 4F systems and face the common issues of 4F systems (discussed above), whereas PhotoFourier is faster, more flexible, and more efficient for accelerating conventional CNNs.

Most on-chip photonic neural network accelerators \cite{shiflett2020pixel, shiflett2021albireo, DEAPCNN2019, jiaqionchipfft, onchip_ONN2018, pcnna, liu2019holylight,zokaee2020lightbulb, 2017mziacc, crosslight} proposed so far are based on MZIs and MRRs. They are typically designed to accelerate dot products and some require IM2COL transformation to compute convolutions. They share similar architecture with other General Matrix Multiplication (GEMM) based accelerators like systolic arrays or compute-in-memory arrays, but can operate faster and utilize wavelength-division multiplexing for extra parallelization. 
Unlike these approaches, PhotoFourier has a fundamentally different architecture by leveraging the complexity reduction of ``free" Fourier transform of Fourier optics to deliver large throughput gains with fewer photonic components (which do not scale well with technology).


\section{Conclusion}
In this paper, we present PhotoFourier, a JTC-based on-chip photonic neural network accelerator. We propose an algorithm to compute 2D convolutions with 1D convolutions that can be implemented using the 1D on-chip lenses. We also propose temporal accumulation to improve the accuracy and power efficiency of the system. Besides, we provide a detailed analysis of how to determine optimal design parameters for a JTC-based CNN accelerator including dataflow, parallelization scheme, and the number of waveguides and compute units. 
Compared to uncompressed photonic neural network accelerators, EDP of PhotoFourier-CG is $28\times$ better compared to Albireo-c, $532\times$ better compared to Holylight-m and $704\times$ better compared to DEAP-CNN.
There are still many remaining challenges for on-chip photonic neural network accelerators, which include the relatively large size and limited flexibility of optical components, manufacturing variations of photonics, data movement cost, and neural network architectures and training methods for photonic execution. We plan to address these issues in our future work, along with a large-scale experimental demonstration of PhotoFourier. 


\clearpage
\bibliographystyle{IEEEtranS}
\bibliography{main}

\end{document}